\documentclass{article}
\usepackage{spconf,amsmath,graphicx}
\usepackage{amssymb}
\usepackage{makecell}
\usepackage{pgfplots}
\usepackage{bm}
\usepackage{hyphenat}
\usepackage{pgfplotstable}
\usepackage{float}

\DeclareMathOperator*{\grad}{\bm \nabla\!}
\DeclareMathOperator*{\gradT}{\bm \nabla^\top\!}
\DeclareMathOperator*{\grads}{\bm \nabla_{\!\sigma}\!}
\DeclareMathOperator*{\gradsT}{\bm \nabla_{\!\sigma}^\top\!}
\DeclareMathOperator*{\gradsg}{\bm \nabla_{\!\sigma,\gamma}\!}
\DeclareMathOperator*{\gradsgT}{\bm \nabla_{\!\sigma,\gamma}^\top\!}

 \pgfplotstableset{
     makebold/.style = {
         postproc cell content/.append style={/pgfplots/table/@cell 
         content/.add={$\bf}{$},},
     },
 }
 
\title{Learning Integrodifferential Models for Image Denoising}
\name{Tobias Alt and Joachim Weickert
      \thanks{This work has received funding from the European Research Council 
              (ERC) under the European Union's Horizon 2020 research and 
              innovation programme (grant agreement no. 741215, ERC Advanced 
              Grant INCOVID).}
     }

\address{Mathematical Image Analysis Group,
         Faculty of Mathematics and Computer Science,\\
         Campus E1.7, Saarland University, 66041 Saarbr\"ucken, Germany.\\
         \{alt, weickert\}@mia.uni-saarland.de}

\begin{document}
\ninept
\maketitle
\begin{abstract}
We introduce an integrodifferential extension of the edge-enhancing 
anisotropic diffusion model for image denoising. By accumulating weighted 
structural information on multiple scales, our model is the first to 
create anisotropy through multiscale integration.
It follows the philosophy of combining the advantages
of model-based and data-driven approaches within compact, insightful, and
mathematically well-founded models with improved performance.
We explore trained results of scale-adaptive weighting and 
contrast parameters to obtain an explicit modelling by smooth functions. This 
leads to a transparent model with only three parameters, without significantly 
decreasing its denoising performance. Experiments demonstrate that it 
outperforms its diffusion-based predecessors. We show that both multiscale 
information and anisotropy are crucial for its success.
\end{abstract}
%


\section{Introduction}\label{sec:intro}
Since three decades, models based on nonlinear partial differential 
equations (PDEs) have been contributing to the progress in image analysis;
see e.g.~\cite{PM90,We97,AK06,LBM19} and the references therein.
PDE models are compact, transparent, and involve only very
few parameters. Moreover, their strict mathematical foundations allow to
establish valuable theoretical guarantees. Well-posedness, for instance,
includes also stability w.r.t.~perturbations of the input data.

More recently, deep learning approaches have been dominating the field 
\cite{GBC16}. Their huge amount of hidden model parameters allow 
a strong adaptation to the training data, which is the reason for their 
excellent performance. On the other hand, these approaches still lack 
transparency and mathematical foundations. They often suffer from 
well-posedness problems such as a high sensitivity to adversarial attacks 
\cite{GSS15}.

In our paper, we are aiming at the best of two worlds: We benefit from 
the compactness and mathematical foundations of PDE-inspired modelling,
while we improve its performance by learning a small set of parameters.
In contrast to most other approaches, however, we compress the model
further by understanding the functional dependence of these parameters.
In other words: {\em We drive learning-based modelling to the extreme
by aiming at the most compact and insightful model.} 
While this philosophy is fairly general, it is best explained by
studying an exemplary application. In this paper, we focus on image 
denoising with edge-enhancing anisotropic diffusion (EED) \cite{We94e}.


\subsection{Our Contribution}
We propose a multiscale extension of EED based on a transparent 
integrodifferential formulation. Our integrodifferential anisotropic diffusion 
(IAD) model adapts to image structures by combining edge information from 
multiple scales and steering the diffusion process accordingly. Given a set of 
images corrupted by Gaussian noise, we learn scale-adaptive parameters of the 
diffusion process by optimising the denoising performance. Afterwards, we 
explore the trained parameters and reduce them by explicitly modelling the 
parameter dynamics over the scales and the noise level. We show that this is 
possible without decreasing the denoising performance significantly. The 
resulting IAD model outperforms both its counterpart of integrodifferential 
isotropic diffusion (IID), as well as EED, while only requiring one additional 
parameter. This shows that both anisotropy and multiscale information are vital 
for the success of the model. 


\subsection{Related Work}
While PDEs are omnipresent in image analysis, models that explicitly involve 
the more general integrodifferential equations are surprisingly rare. They 
are considered occasionally for image decompositions \cite{AT11}, 
nonlocal generalisations of PDE evolutions \cite{GO08,CWS15}, and in connection
with fractional calculus for linear image processing \cite{CF03}. These 
works are only very mildly related to our paper.

Since our IAD model sums up contributions from multiple scales,
it has some conceptual similarities to wavelet shrinkage.
Didas and Weickert~\cite{DW07} relate wavelet shrinkage to 
nonlinear diffusion, but only for the isotropic case. As each scale is 
treated independently, anisotropy cannot be achieved. Welk et al.~\cite{WWS06} 
present an anisotropic diffusion method based on wavelet shrinkage, but 
only on the finest scale. We combine the best of these two ideas into a novel 
model: Our IAD model combines structural image information over multiple scales 
to create an anisotropic diffusion process.

Integrodifferential models for nonlinear diffusion predominantly involve 
models with Gaussian-smoothed derivatives. Most of these models 
\cite{CLMC92,We94e,NRFV97,SchW98} have been proposed as regularisations of 
the Perona--Malik filter \cite{PM90}. For enhancing coherent structures, 
one also considers a smoothed structure tensor \cite{FG87} that captures 
directional information to steer the diffusion process accordingly 
\cite{We97}. However, all of these models only consider Gaussian smoothing 
on a fixed scale and do not incorporate an integration over multiple scales.

Large-scale trainable models involving PDEs
have recently become very successful. Chen and Pock \cite{CP16} train flux 
functions and derivative filters of a diffusion-inspired model to obtain 
exceptional denoising results. Other authors only train nonlinearities of 
models \cite{SS14,BL20,AW20} or learn PDEs directly with 
symbolic approaches \cite{Sch17,LLD19,RMMW20}.
Instead of focusing on denoising performance, we want to obtain insights into 
the scale behaviour of our model. In contrast to \cite{CP16}, we have 
stability guarantees in the Euclidean norm and require very few parameters 
that allow insights into their functional dependency. This 
helps us to obtain a model which is transparent and outperforms its
diffusion-based predecessors.


\subsection{Organisation of the Paper}
In Section \ref{sec:review}, we review nonlinear diffusion. This serves as 
a basis for our novel IAD model that is introduced in Section \ref{sec:ours}. 
In several experiments in Section \ref{sec:experiments}, we reduce its 
parameters and compare it to other diffusion filters. Finally, 
we present our conclusions and give an outlook on future work in Section 
\ref{sec:conclusion}. 


\section{Review of Nonlinear Diffusion}\label{sec:review}

\subsection{Isotropic Case: \hspace{0mm} The Perona--Malik Model}
Let us consider some rectangular image domain $\Omega \subset \mathbb{R}^2$.
The classical nonlinear diffusion model of Perona and Malik \cite{PM90} 
embeds a greyscale image $f(x,y): \Omega \rightarrow 
\mathbb{R}$ into a family of simplified versions 
$\{u(x,y,t) \,|\, t \ge 0\}$ by solving the diffusion equation 
\begin{equation} \label{eq:pm}
  \partial_t u  =  \gradT\!\left(g_\lambda(|\!\grad u|^2) \bm \grad u\right),
\end{equation}
where $\grad = (\partial_x, \partial_y)^\top$ denotes the spatial gradient 
operator and $|\,.\,|$ the Euclidean norm. The process is initialised with 
$u(x,y,0)\!=\!f(x,y)$, and on
the domain boundary $\partial \Omega$ one uses reflecting boundary 
conditions. Larger diffusion times $t$ yield stronger diffused 
representations.

To enable a diffusion process that respects edges, Perona and Malik 
use $\left|\grad u\right|^2$ as an edge indicator and employ a decreasing
scalar-valued diffusivity function in $\left|\grad u\right|^2$, e.g. 
\begin{equation} \label{eq:diffusivity}
g_\lambda(\left|\grad u\right|^2) = \exp\!\left(-\frac{\left|\grad u\right|^2}
{2 \lambda^2}\right)
\end{equation}
with a contrast parameter $\lambda>0$.
For large values of $\left|\grad u\right|^2$, the diffusivity becomes very 
small, such that the edge is preserved. However, while a scalar-valued 
diffusivity function can respect the {\em location} of an edge, it is unable to
represent its {\em direction}. Therefore, we regard the Perona--Malik 
model as {\em isotropic} in the physical sense of diffusion. At noisy
edges, the Perona--Malik model reduces the diffusivity and preserves the
edge without denoising it.


\subsection{Anisotropic Case: \hspace{0mm} The EED Model}

The edge-enhancing anisotropic diffusion (EED) model \cite{We94e} 
avoids this drawback of the Perona--Malik model by replacing its 
scalar-valued diffusivity by a matrix-valued diffusion
tensor. This positive semidefinite $2 \times 2$ matrix 
is designed in such a way that smoothing along edges is encouraged, while 
smoothing across them is inhibited. Its normalised eigenvectors $\bm{v}_1$
and $\bm{v}_2$ are chosen parallel and perpendicular to a Gaussian-smoothed
image gradient $\grads u$, where $\sigma$ denotes the standard deviation of
the Gaussian. The corresponding eigenvalues $\lambda_1$, $\lambda_2$ are
modelled as 
\begin{equation}
 \lambda_1(|\!\grads u|^2) =g_\lambda(|\!\grads u|^2), \qquad
 \lambda_2(|\!\grads u|^2) =1. 
\end{equation}
This uniquely defines the diffusion tensor as
\begin{equation}
  \bm{D}_\lambda(\grads u) 
  = g_{\lambda}\!\left(\left|\grads u\right|^2\right) \, 
    \bm v_1 \bm v_1^\top + 1 \, \bm v_2 \bm v_2^\top.
\end{equation}
Note that anisotropy results form the fact that the eigenvector 
$\,\bm{v}_1 \parallel \grads u\,$ of the diffusion tensor in general 
does not coincide with the 
image gradient in the product $\,\bm D_{\lambda} (\grads u)  \grad u$. 
Otherwise, we obtain $\,\bm D_{\lambda} (\grads u) \grad u=
g_{\lambda}\!\left(\left|\grads u\right|^2\right) \grad u$, which shows
that the process becomes isotropic with a scalar-valued diffusivity. 
This is also the case for $\sigma \rightarrow 0$.


\subsection{A Useful Reformulation of EED}
To prepare ourselves for our novel integrodifferential model, we
reformulate the EED model with the structure tensor \cite{FG87}
\begin{equation}
\bm J = \grads u \gradsT u.
\end{equation}
By plugging in, we see that the matrix $\bm{J}$ has 
eigenvectors $\bm{v}_1 \parallel \grads u$ and $\bm{v}_2 \perp \grads u$ 
with eigenvalues $\nu_1=\left|\grads u\right|^2$ and $\nu_2=0$.

Expressing the diffusivity function $g_{\lambda}$ from (\ref{eq:diffusivity}) 
in terms of a power series, we can generalise it to matrix-valued arguments 
such as $\bm J$. Then $g_{\lambda}(\bm J)$ is a matrix as well. It has 
the same eigenvectors $\bm{v}_1$, $\bm{v}_2$ as $\bm J$, and its
eigenvalues satisfy
$\lambda_1 = g_\lambda(\nu_1) = g_\lambda(|\!\grads u|^2)$ and
$\lambda_2 = g_\lambda(\nu_2) = g_\lambda(0) = 1$.
Thus, we have $\bm{D}_\lambda(\grads u)=g_\lambda(\bm{J})$ and can write
the EED evolution as
\begin{equation}
  \partial_t u  =  \gradT\!\left(g_\lambda(\bm{J})
  \bm \grad u\right).
\end{equation}
It closely resembles the Perona--Malik equation
(\ref{eq:pm}). Indeed, by replacing the matrix-valued tensor product
$\bm{J}=\grads u \gradsT u$ by the scalar-valued inner product
$\gradsT u \grads u =\left|\grads u\right|^2$ and taking $\sigma~\to~0$,
we end up with the Perona--Malik model. Thus, there is an elegant
connection between isotropic and anisotropic models.

The choice of the scale $\sigma$ is crucial for the denoising 
performance of EED. A single scale may be too coarse for fine-scale details, 
while at the same time not capturing large-scale structures. We will 
see that by accumulating information over multiple scales within the 
structure tensor, we obtain a more useful representation of image 
information which again leads to improved denoising results.


\section{Integrodifferential Diffusion}\label{sec:ours}

\subsection{Continuous Model}

To account for multiscale information within the structure tensor, we 
introduce the following \emph{integrodifferential anisotropic diffusion 
(IAD)} model:
\begin{equation}
  \partial_t u = \int_{0}^{\infty}
  \gradsgT \Big(g_{\lambda}\!\left(
      \bm J_\gamma \right) \, \gradsg u \Big) \, d\sigma,
\end{equation}
where the \emph{multiscale structure tensor}
\begin{equation}
\bm J_\gamma = \int_{0}^{\infty} \!
 \gradsg u \, \gradsgT u \; d\sigma
\end{equation} 
accumulates structural 
information over all smoothing scales $\sigma$. This integration generates 
anisotropy on each scale, since the eigenvectors of $\bm J_\gamma$ usually 
are not parallel to $\gradsg u$.

To adapt the contribution of each scale, we introduce an additional weight 
parameter $\gamma(\sigma)$ in the smoothed gradient, i.e. $\gradsg u = \grad 
\left(\gamma(\sigma) K_\sigma \ast u\right)$. Here, $K_\sigma$ is a Gaussian of 
standard deviation $\sigma$. Its normalisation is subsumed by $\gamma(\sigma)$.
In contrast to other models employing an outer smoothing with a fixed weighting 
\cite{WB02,RM07}, the IAD model allows to attenuate scales if they do 
not offer useful information. This weighting, along with the multiscale 
structure tensor, is the key to the success of the IAD model.

The diffusion tensor $g_\lambda \! \left(\bm J_\gamma\right)$ inherits 
its eigenvectors from $\bm{J}_\gamma$, and its eigenvalues are given by 
$g_\lambda(\mu_1)$ and $g_\lambda(\mu_2)$ where $\mu_1$,  $\mu_2$ are 
the eigenvalues of $\bm{J}_\gamma$.
Similar diffusion tensors are considered e.g.~in \cite{WB02,PW15a}.
Moreover, we make the contrast parameter $\lambda(\sigma)$ of the 
diffusivity scale-adaptive. This individually steers the balance between 
smoothing and edge enhancement on each scale.

An isotropic counterpart of the IAD model, which we call 
\emph{integrodifferential isotropic diffusion (IID)}, arises directly by 
switching the order of transposition within the structure tensor. In 
this case,
\begin{equation}
\int_{0}^{\infty} \gradsgT u \gradsg u \: d\sigma 
= \int_{0}^{\infty}\left|\gradsg u\right|^2 
\, d\sigma
\end{equation}
is no structure tensor any more, but a multiscale gradient magnitude.
In our experiments, we will use the IID model for comparisons.


\subsection{Trainable Discrete Model}

In a practical setting, we need a discrete version of the continuous IAD model. 
To this end, we employ an explicit finite difference scheme: We discretise the 
temporal derivative by a forward difference with time step size $\tau$, 
and the right-hand side by the nonnegativity discretisation from 
\cite{We97}. Additionally, we select a set of $N$ discrete 
scales $\sigma_1,\dots,\sigma_N$ according to an exponential sampling. 
This yields discrete weights $\gamma_i=\gamma(\sigma_i)$ and 
contrast parameters $\lambda_i=\lambda(\sigma_i)$, which we obtain by 
training.
We iterate the method for a small number of $K$ explicit steps. This
yields an approximation of the continuous model for a diffusion time of
$T=K\tau$. Iterating the IAD model allows to capture a nonlinear evolution.

As usual, our explicit scheme must satisfy a time step size restriction that 
is governed by the spectral radius of the iteration matrix; 
see e.g.~\cite{DWB09} for more details. 
Then the following two important properties are easy to establish: 
\begin{itemize}
\item The discrete IAD model is stable in the Euclidean norm, i.e.~its
      Euclidean norm is nonincreasing in each step.
\item Since the algorithm consists of a concatenation of continuous
      function evaluations, it constitutes a well-posed discrete evolution.
      In particular, this implies that the output depends continuously on the 
      input data.
\end{itemize}


\subsection{Learning Framework} 

To train the $2N$ parameters $\gamma_i$ and $\lambda_i$ of the discrete model, 
we consider a training set of  
$200$ grey value images of size $256\!\times\!256$ with grey value range $[0, 
255]$. We crop them from the BSDS500 dataset \cite{AMFM11} and corrupt them 
with additive Gaussian noise of standard deviation $s$. The resulting grey 
values are not cut off if they exceed the original grey value range to preserve 
the Gaussian statistics of the noise. A disjoint test set of $100$ images is 
generated accordingly.

We train the model with a sufficient number of explicit 
steps by minimizing the average mean square error over the training set. 
Moreover, we always choose $\tau$ in such a way that it is stable and allows 
for a suitable diffusion time. We found that $K=10$ steps are already 
sufficient for noise levels up to $s = 60$.


\section{Experiments}\label{sec:experiments}

\subsection{Finding Scale-adaptive Parameter Functions}
To gain insights into the behaviour of the IAD model, we develop 
smooth relations for the discrete trained parameters in terms of the scale 
$\sigma$ and the noise standard deviation $s$.

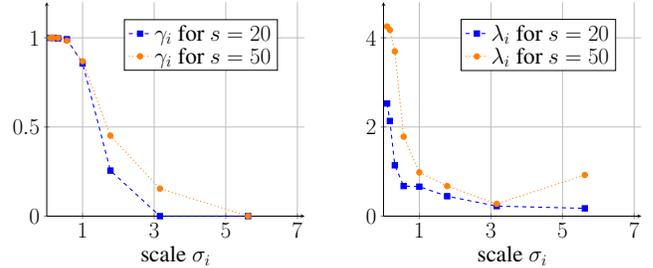
\begin{figure}
  \centering
  \begin{tikzpicture}[scale=0.5]

\begin{axis}
[
samples=500, 
font=\LARGE,
domain=0:8.1, 
axis lines=left,
ymin = 0, ymax = 1.2,
xmin = 0, xmax = 7.2,
xtick = {1,3,5,7},
ytick = {0, 0.5, 1},
xlabel={scale $\sigma_i$},
xlabel near ticks,
ylabel near ticks,
grid = major,
legend entries={$\gamma_i$ for $s=20$,
                $\gamma_i$ for $s=50$}, 
legend cell align=left,
legend style={at={(0.3, 0.8)}, anchor = west}
]

\addplot+[thick, dashed, mark options={solid}, blue, mark = square*,
          restrict expr to domain={\thisrow{noise}}{20:20}]
table[x = scale, y expr = \thisrow{gamma}/0.07814744] 
{figures/fig_param_inspection/params.dat};

\addplot+[thick, dotted, mark options={solid}, orange, mark = *,
          restrict expr to domain={\thisrow{noise}}{50:50}]
table[x = scale, y expr = \thisrow{gamma}/0.05434512] 
{figures/fig_param_inspection/params.dat};


\end{axis}

\end{tikzpicture}
  \hspace{5mm}
  \begin{tikzpicture}[scale=0.5]

\begin{axis}
[
samples=500, 
font=\LARGE,
domain=0:8.1, 
axis lines=left,
ymin = 0, ymax = 4.8,
xmin = 0, xmax = 7.2,
xtick = {1,3,5,7},
ytick = {0, 2, 4},
xlabel={scale $\sigma_i$},
xlabel near ticks,
ylabel near ticks,
grid = major,
legend entries={$\lambda_i$ for $s=20$,
                $\lambda_i$ for $s=50$},
legend cell align=left,
legend style={at={(0.3, 0.8)}, anchor = west}
]

\addplot+[thick, dashed, mark options={solid}, blue, mark = square*,
          restrict expr to domain={\thisrow{noise}}{20:20}]
table[x = scale, y expr = \thisrow{lambda}*255] 
{figures/fig_param_inspection/params.dat};

\addplot+[thick, dotted, mark options={solid}, orange, mark = *,
          restrict expr to domain={\thisrow{noise}}{50:50}]
table[x = scale, y expr = \thisrow{lambda}*255] 
{figures/fig_param_inspection/params.dat};


\end{axis}

\end{tikzpicture}
  \caption{Learned weight and contrast parameters $\gamma_i, \lambda_i$  
  for two different noise standard deviations $s$.
  \label{fig:inspect_param}}
\end{figure}

To this end, we train the full model for $K=10$ explicit steps and $N=8$ 
discrete scales for varying noise standard deviations 
$s\in\{10,20,\dots,60\}$. 
To ensure that the parameters $\gamma_i$ and $\lambda_i$ vary smoothly over the 
scales, we add a small regularisation to the optimisation loss. It penalises 
variations of $\gamma$ and $\lambda$ over the scales in the squared euclidean 
norm. We present the resulting parameter dynamics over the scales in 
Fig.~\ref{fig:inspect_param} for two noise levels each. 

For the weight parameters $\gamma_i$, we find that the importance of the 
structural information decreases with the scale. Note that we normalised the 
weight parameters such that $\gamma_1 = 1$, as a global rescaling can be 
subsumed within the time step size $\tau$. The contribution of smoothing scales 
with $\sigma>5$ is almost non-existent. This is in accordance with results for 
scale-adaptive wavelet shrinkage \cite{AW20}, where useful shrinkage is only 
performed on the finer scales. 
Furthermore, we observe that larger noise levels $s$ lead to a higher 
weighting of larger scales: As the images are corrupted more strongly, larger 
scales are needed to extract useful structural information. 
Based on Fig.~\ref{fig:inspect_param}, we approximate the parameter dynamics of 
the weight function $\gamma(\sigma, s)$ by
\begin{equation}
  \gamma\!\left(\sigma, s\right) = 
  \exp\!\left(-\frac{\alpha\sigma^2}{\sqrt{s}}\right),
\end{equation} 
with a learned scalar $\alpha>0$ that steers the decay.

Similarly, we also see that the nonlinear response of the diffusivity function 
$g_\lambda$ decreases with larger scales, however not as rapidly as the weights 
$\gamma$. As noise is attenuated on coarser scales and structural information 
dominates, 
the balance between structure preservation and noise removal is shifted 
accordingly by decreasing $\lambda$. Outliers for large scales can be 
ignored, as $\gamma \rightarrow 0$ dampens the influence of $\lambda$. 
When increasing the noise level, the contrast parameters scale linearly, a 
relation which has also been observed for wavelet shrinkage \cite{HS08}. 
Therefore, we model the contrast parameters as 
\begin{equation}
  \lambda\!\left(\sigma, s\right) = \frac{\lambda_0s}{1 + \beta 
  \sigma^2},
\end{equation} 
with learned scalars $\beta, \lambda_0 > 0$, where $\beta$ determines the 
decay of the curve and $\lambda_0$ its value at $\sigma=0$.
Through these insights, we have effectively reduced the parameter set to only 
three trainable parameters $\alpha$, $\beta$, $\lambda_0$, one more than for 
EED. 

In an ablation study, we now show that we hardly sacrifice any performance 
for the sake of parameter reduction. In a first step, we train the full 
model for 
$K=10$ explicit steps, $N=8$ scales, and noise levels 
$s\in\{10,20,\dots,60\}$ without any smoothness regularisation. With two 
parameters per scale and noise level, this amounts to $96$ parameters. 
Afterwards, we train the parameters $\alpha$, $\beta$, and $\lambda_0$ of the 
reduced model for the same configuration jointly for all noise levels. We 
obtain $\alpha=1.64$, $\beta=2.46$ and $\lambda_0=1.47$. 

On average, the fully 
trained model with $96$ parameters performs only $0.07$~dB better than the 
reduced model with three parameters in terms of peak signal-to-noise ratio
(PSNR). This marginal performance difference shows that the parameter 
relations which we have modelled represent the true parameter dynamics well.

\setlength{\tabcolsep}{2pt}
\begin{figure}[!t]
  \centering
  \begin{tabular}{cc}
  \includegraphics[width=0.48\linewidth]
   {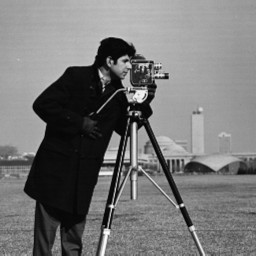}
  &\includegraphics[width=0.48\linewidth]
   {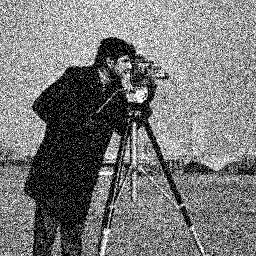}
  \\
    ground truth
  & noisy, \hspace{0mm} $s=50$
  \\[2mm]
  \includegraphics[width=0.48\linewidth]
    {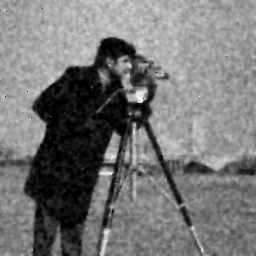}
  &\includegraphics[width=0.48\linewidth]
   {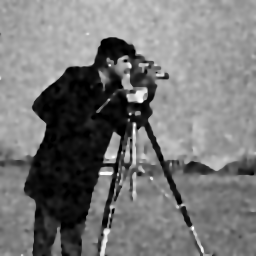}
  \\
    \makecell{PM,  \hspace{0mm} PSNR $=24.21$ dB} 
  & \makecell{EED, \hspace{0mm} PSNR $=25.60$ dB}
  \\[2mm]
  \includegraphics[width=0.48\linewidth]
   {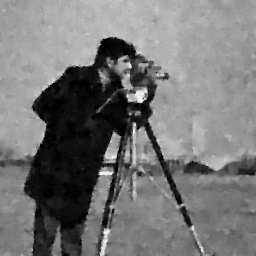}
  &\includegraphics[width=0.48\linewidth]
   {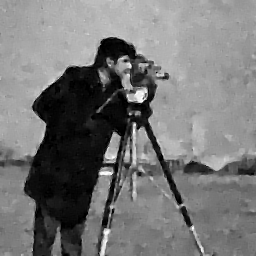}
  \\
    \makecell{IID,  \hspace{0mm} PSNR $=25.63$ dB} 
  & \makecell{IAD,  \hspace{0mm} PSNR $=\bm{26.02}$ dB}
  \end{tabular}
  \caption{Qualitative comparison of the denoising performance on the 
  \emph{cameraman} image. The multiscale information of IID and IAD helps to 
  adapt smoothing to different scales, and the anisotropy of EED and IAD 
  allows for directional smoothing along edges. \label{fig:image_comparison}}
\end{figure}

\subsection{Comparison to Other Diffusion Filters}
We compare the reduced model to the isotropic Perona--Malik model (PM) 
\cite{PM90}, as well as EED \cite{We94e}. Additionally, we consider IID as the 
isotropic variant of the IAD model. This four-way comparison is designed in an 
ablative way to show that both multiscale modelling and anisotropy are 
crucial for the success of the IAD model: Multiscale 
information is not considered for EED and PM, while anisotropy is not used for 
PM and IID. We explicitly compare only to other diffusion-based methods, as it 
is our intention to transparently improve those compact and insightful models 
rather than to produce state-of-the-art results.  

To ensure a fair comparison, all models are trained with the exponential 
Perona--Malik diffusivity. 
For PM and EED, we optimise the 
parameters for each noise level individually. However, for IID and IAD, we 
respectively use the reduced parameters over the full noise range.

Fig.~\ref{fig:image_comparison} shows representative denoising results on 
the \emph{cameraman} image for $s=50$. For the Perona--Malik model, the 
optimal contrast parameter is too small to remove extreme noise outliers 
but too large to preserve edges. EED yields sharp edges in the 
foreground. However, due to the single smoothing scale, it is not 
able to denoise the background efficiently, where a much larger smoothing scale 
would be appropriate. The IID and IAD models do not suffer 
from this effect as the multiscale information can identify structures on 
all scales. However, IID suffers from remaining noise around edges. 
This is an intrinsic drawback of the isotropic model which detects only the 
location of an edge, but not its orientation. Finally, IAD combines edge 
enhancement together with efficient denoising in homogeneous regions. 

\begin{table}
 \setlength{\tabcolsep}{6pt}
 \renewcommand{\arraystretch}{1.1}
 \centering
 \begin{filecontents*}{figures/fig_quality_comparison/results.dat}
stddev  PM      EED     IID     IAD
10      32.00   32.15   32.10   32.41
20      28.02   28.33   28.32   28.64
30      25.94   26.37   26.43   26.74
40      24.55   25.11   25.25   25.52
50      23.56   24.24   24.41   24.61
60      22.83   23.55   23.78   23.81
\end{filecontents*}
\pgfplotstabletypeset[
 col sep=space,
 string type,
 every head row/.style={after row=\hline},
 columns/stddev/.style={column name=$s$, column type=c|},
 columns/PM/.style={column name=PM, column type=c},
 columns/EED/.style={column name=EED, column type=c},
 columns/IID/.style={column name=IID, column type=c},
 columns/IAD/.style={column name=IAD, column type=c, makebold},
 ]{figures/fig_quality_comparison/results.dat}
  \caption{Average PSNR on the test set of the four models considered. Higher 
  values indicate better quality. \label{tab:comparison}}
\end{table}

Lastly, we compare the performance of the four models on different noise 
levels. Table \ref{tab:comparison} shows the average PSNR on the test 
set. Surprisingly, already the IID model is able to beat EED for Gaussian 
noise with standard deviation $s \ge 30$. With its additional anisotropy, 
the IAD model consistently outperforms its competitors. With higher amounts of 
noise, IID and IAD increase the gap to their PDE-based predecessors, 
indicating that multiscale information becomes increasingly important 
with noise.


\section{Conclusions}\label{sec:conclusion}
We have shown that one can elegantly introduce multiscale information within 
an anisotropic diffusion process by considering integro\-differential models.  
Learning was involved to uncover the optimal scale dynamics of the resulting 
model.

Apart from their evident merits of improving anisotropic diffusion methods, 
the findings in our paper are of more fundamental nature in two aspects:
Firstly, we have seen that integrodifferential equations are hitherto 
hardly explored but very promising extensions of differential 
equations. They are not only more general, but also allow new 
possibilities for data adaptation, e.g.~by anisotropy through 
multiscale integration.
Secondly, our paper illustrates the potential of learning with
maximal model reduction. Such approaches can benefit from the
best of two worlds: the transparency and mathematical foundation of
model-based techniques and performance improvements by data-driven,
learning-based strategies.

Showing these advantages in a more general context is part of our
future research. We are exploring applications beyond denoising, and 
we will also study integrodifferential extensions of other important 
PDE models.


\newpage
\bibliographystyle{IEEEbib}
\bibliography{myrefs}

\end{document}